\documentclass{svproc}
\usepackage{graphicx}
\usepackage{bm}
\usepackage{amsmath}

\usepackage{url}

\begin{document}
\mainmatter              
\title{Interplay of Coulomb repulsion and spin-orbit coupling in superconducting 3D quadratic band touching Luttinger semimetals}
\titlerunning{Superconductivity in Luttinger semimetals}  
%
\author{Serguei Tchoumakov\inst{1} \and Louis J. Godbout\inst{1} \and William Witczak-Krempa\inst{1,2,3}}
\authorrunning{Serguei Tchoumakov et al.} 
%
\tocauthor{Serguei Tchoumakov, Louis J. Godbout, and William Witczak-Krempa}
\institute{D\'epartement de Physique, Universit\'e de Montr\'eal, Montr\'eal, Qu\'ebec, H3C 3J7, Canada\\
\email{serguei.tchoumakov@umontreal.ca}
\and
Centre de Recherches Math\'ematiques, Universit\'e de Montr\'eal; P.O. Box 6128, Centre-ville Station; Montr\'eal (Qu\'ebec), H3C 3J7, Canada
\and
Regroupement Qu\'eb\'ecois sur les Mat\'eriaux de Pointe (RQMP)
}

\maketitle              

\begin{abstract}
We investigate the superconductivity of 3D Luttinger semimetals, such as YPtBi, where Cooper pairs are constituted of spin-$3/2$ quasiparticles. Various pairing mechanisms have already been considered for these semimetals, such as from polar phonons modes, and in this work we explore pairing from the screened electron-electron Coulomb repulsion.

In these materials, the small Fermi energy and the spin-orbit coupling strongly influence how charge fluctuations can mediate pairing. We find the superconducting critical temperature as a function of doping for an $s-$wave order parameter, and determine its sensitivity to changes in the dielectric permittivity. Also, we discuss how order parameters other than $s-$wave may lead to a larger critical temperature, due to spin-orbit coupling.
\end{abstract}
\section{Introduction}
In regular metals, Coulomb repulsion is commonly believed to compete against the superconducting pairing between electrons. For example, in the electron-phonon mechanism of superconductivity with electron-phonon coupling $g$, the critical temperature below which an electron gas becomes superconducting is \cite{macmillan}
\begin{equation}
  T_c = \frac{\langle \omega \rangle}{1.2} \exp\left(- \frac{1.04(1+\lambda)}{\lambda - \mu^* (1+0.62\lambda)}\right),
  \label{eq:macmillan}
\end{equation}
where $\lambda = 2 \int_{0}^{\infty}d\omega g^2D(\omega)/\omega$ is the coupling constant, $\langle \omega \rangle = 2 \int_{0}^{\infty}d\omega g^2D(\omega)$ is the averaged phonon frequency and $\mu^*$ is the Coulomb pseudo-potential. $D(\omega)$ is the phonon density of states. Eq.~(\ref{eq:macmillan}) informs us that increasing the electronic Coulomb repulsion exponentially decreases the critical temperature. In a normal electron gas, this strength is measured by the Wigner-Seitz radius $r_s \approx 2 m e^2/\epsilon^* k_F$ where $m$ is the band mass, $e$ the electronic charge, $\epsilon^*$ the background dielectric constant and $k_F$ is the Fermi wavevector. The superconductivity in semiconductors is often attributed to the electron-phonon coupling. However, for some materials such as SrTiO$_3$ and bismuth-based half-Heuslers, like YPtBi, the importance of the electron-phonon coupling in superconductivity is yet unresolved. In SrTiO$_3$ it was even proposed that superconductivity may come from the electron-electron repulsion \cite{takada,ruhman}. The qualitative explanation does not only rely on the Kohn-Luttinger mechanism \cite{kohnluttinger} but also on the contribution of plasmons to screening \cite{takada}. The effective attraction between electrons is a consequence of the screening of the Coulomb potential, with a dielectric function $\epsilon_{}(\omega,{\bf q})$ that is computed in the random phase approximation
\begin{equation}
  \epsilon_{\rm RPA}(\omega,{\bf q}) = 1 - V_0({q})\Pi_0(\omega,{\bf q}),
\end{equation}
with $V_0({q}) = 4\pi e^2/(\epsilon^*q^2)$ the bare Coulomb potential and $\Pi_{0}(\omega,{\bf q})$ the bare electron polarisability. The dielectric function $\epsilon(\omega, {\bf q})$ depends on the system under study and has a role similar to the density of states of phonons, $D(\omega)$ that appears in Eq.~(\ref{eq:macmillan}). In Ref.~\cite{scus} we use a variational approach similar to that in \cite{allen}
to show how the critical temperature depends on each component $(\omega,{\bf q})$ of the dielectric function, as we discuss further below.

\begin{figure} [t]
    \centering
    \includegraphics[width = \textwidth]{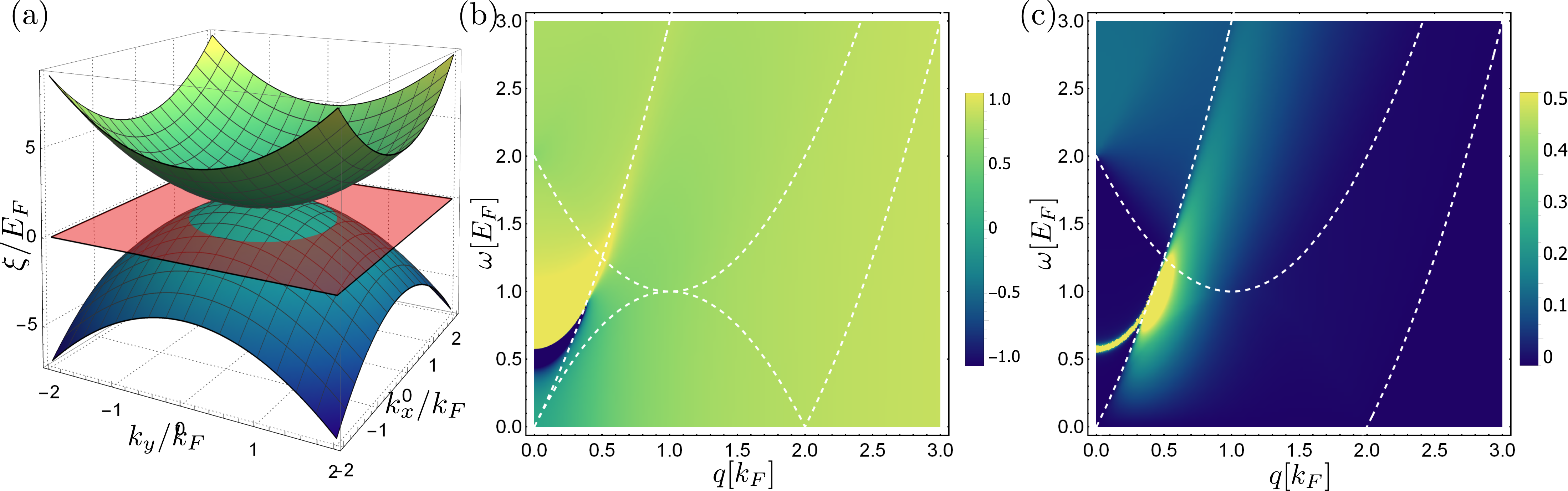}
    \caption{(a) Band-structure of a Luttinger semimetal, the red plane is at the Fermi level. The upper and lower bands are doubly degenerate. (b-c) The (b) real and (c) imaginary parts of the inverse dielectric permittivity, $1/\epsilon(\omega,q)$, for $r_s = 0.5$ as a function of wavevectors, $q$, and frequencies, $\omega$. The white dashed lines are the branches of the particle-hole continuum.}
    \label{fig:screen}
\end{figure}

This mechanism for SrTiO$_3$ however does not directly apply to bismuth-based half-Heusler materials, such as YPtBi, where the band structure is not well approximated by the free Hamiltonian $H_{\rm N}({\bf k}) = \hbar^2 k^2/(2m) - \mu$ but also includes strong spin-orbit coupling. It is a candidate Luttinger semimetal with Hamiltonian~\cite{luttinger}
\begin{equation}
  \hat{H}_0({\bf k}) = \frac{\hbar^2}{2m} \left[   - \frac{5}{4} {\bf k}^2  + \left({\bf k}\cdot \hat{{\bf J}}\right)^2 \right]  - \mu,
  \label{eq:h0}
\end{equation}
where we introduce the $j = 3/2$ total angular momentum operators $\hat{\bf J} = (\hat{J}_x,\hat{J}_y,\hat{J}_z)$ and the chemical potential $\mu$. This model has inversion, rotational and time-reversal symmetries. The spectrum consists of four bands that meet quadratically at ${\bf k} = 0$ with degenerate lower and upper bands with energies $\pm \hbar^2 k^2/(2m)$ as shown in Fig.~\ref{fig:screen}(a). In the present proceeding we outline our findings regarding screening, quasiparticles and superconductivity in Luttinger semimetals arising from the screened Coulomb repulsion \cite{screself,scus}. We also discuss how the $J = L = S = 1$ order parameter may have a larger critical temperature than in the $s-$wave channel, due to spin-orbit coupling.

\section{Screening and electronic properties of Luttinger semimetals}

We perturb the bare Hamiltonian (\ref{eq:h0}) with the bare Coulomb potential $V_0(q)$
\begin{equation}
  \hat{H}_{\rm int} = \frac{1}{2\mathcal{V}} \sum_{{s_1 s_2{\bf k}_1 {\bf k}_2,{\bf q}\neq 0}} V_0({ q}) \hat{\psi}^{\dagger}_{{\bf k}_1 + {\bf q}s_1} \hat{\psi}^{\dagger}_{{\bf k}_2 - {\bf q}s_2} \hat{\psi}_{{\bf k}_2s_2}\hat{\psi}_{{\bf k}_1s_1},
  \label{eq:hint}
\end{equation}
where $\mathcal{V}$ is the volume of the electron gas and introduce the annihilation operators $\hat{\psi}_{{\bf p}s} = \{\hat{\psi}_{{\bf p},3/2},\hat{\psi}_{{\bf p},1/2},\hat{\psi}_{{\bf p},-1/2},\hat{\psi}_{{\bf p},-3/2}\}$ of the aforementioned $j=3/2$ representation. In the following, we set $\hbar = k_B = 1$ with energies in units of the Fermi energy $E_F$ and wavevectors in units of the Fermi wavevector $k_F$. The amplitude of the Coulomb potential is then given by the Wigner-Seitz radius, $r_s = me^2/(\alpha \epsilon^* k_F)$ with $\alpha \approx 0.51$. In \cite{screself} we computed the bare charge polarisability $\Pi_0(\omega,{\bf q})$ and the self-energy corrections $\Sigma_{\pm}(\omega, {\bf k})$ on the upper ($+$) and lower ($-$) bands. We find that, because of strong spin-orbit coupling, the plasma frequency is diminished compared to a regular quadratic band, and that screening receives important contributions from interband excitations (see Figs.~\ref{fig:screen}(b,c)).


The difference in screening between a Luttinger semimetal and a normal electron gas affects the quasiparticle properties. We find that for Luttinger semimetals the quasi-particle residue $Z_F$ and the first Landau coefficients, $f_{0s}$ and $f_{1s}$, are less affected by the Coulomb potential \cite{screself}. 

\section{Superconductivity in Luttinger semimetals}
\label{sec:sc}

We evaluate the critical temperature of a singlet $s-$wave superconductor using the linear Eliashberg equation \cite{ruhman,takada2}, with account of self-energy corrections,
\begin{equation}
  \lambda(T) \phi_{\sigma_1}(i\omega_{n_1}, k_1) = -T\!\!\sum_{\sigma_2\omega_{n_2}}\!\!\!\int_{0}^{\infty}\!\!\!\!dk_2 \frac{k_2}{k_1}\frac{I_{0\sigma_1\sigma_2}(i\omega_{n_1}, k_1 ; i\omega_{n_2}, k_2) \phi_{\sigma_2}(i\omega_{n_2},k_2)}{(\omega_{n_2}Z_{\sigma_2}(i\omega_{n_2},k_2))^2 + (\xi_{\sigma_2}(k_2) + \chi_{\sigma_2}(i\omega_{n_2}, k_2))^2},
  \label{eq:linelsh}
\end{equation}
where $\phi_{\sigma}$ represents the superconducting order parameter, $\omega_n = (2n+1)\pi T$ are the Matsubara frequencies, $\sigma = \pm$ is the band index, $I_0$ is the angular average of the screened Coulomb potential with spin-orbit corrections and $\Sigma_{\pm}(i\omega_n, k) \equiv \chi_{\pm}(i\omega_n,{k}) + i\omega_n(1 - Z_{\pm}(i\omega_n,k))$ are the self-energy corrections. Note that we have included the pairing order parameter on the upper band ($+$), as it will play an important role. Eq.~(\ref{eq:linelsh}) is an eigenvalue equation where the critical temperature is found for eigenvalues $\lambda(T)$ such that $\lambda(T_c) = 1$.

In this approach, the absence of symmetry in Eq.~(\ref{eq:linelsh}) on parameters ($\sigma, \omega_n, k$) makes its resolution complex and time consuming
. We thus perform the transformation $\phi_{\sigma}(i\omega_n,k) \rightarrow \bar{\phi}_{\sigma}(i\omega_n,k) = k\phi_{\sigma}(i\omega_n,k)/((\omega_{n}Z_{\sigma}(i\omega_{n},k))^2 + (\xi_{\sigma}(k) + \chi_{\sigma}(i\omega_{n}, k))^2)$ to have a symmetric form of Eq.~(\ref{eq:linelsh})
\begin{equation}
  \label{eq:elshsym}
  \rho(T) \bar{\phi} = S\bar{\phi},
\end{equation}
with $S$ a symmetric operator on parameters ($\sigma, \omega_n, k$) and where the critical temperature $T_c$ is obtained for $\rho(T_c) = 0$. One can show that $\rho(T > T_c) < 0$, so $T_c$ is computed from the largest eigenvalue $\rho^{\rm max}$ and, using the variational properties of symmetric matrices, for any test function $\bar{\phi}^{\rm t}$ :
\begin{equation}
  \rho^{\rm max} \geq \rho^{\rm t} = \frac{\bar{\phi}^{\rm t}\cdot S\bar{\phi}^{\rm t}}{\bar{\phi}^{\rm t}\cdot\bar{\phi}^{\rm t}} \Rightarrow T_c \geq T_c^{\rm t},
\end{equation}
with $T_c^{\rm t}$ the critical temperature obtained with the test function.

\begin{figure} [t]
    \centering
    \includegraphics[width = \textwidth]{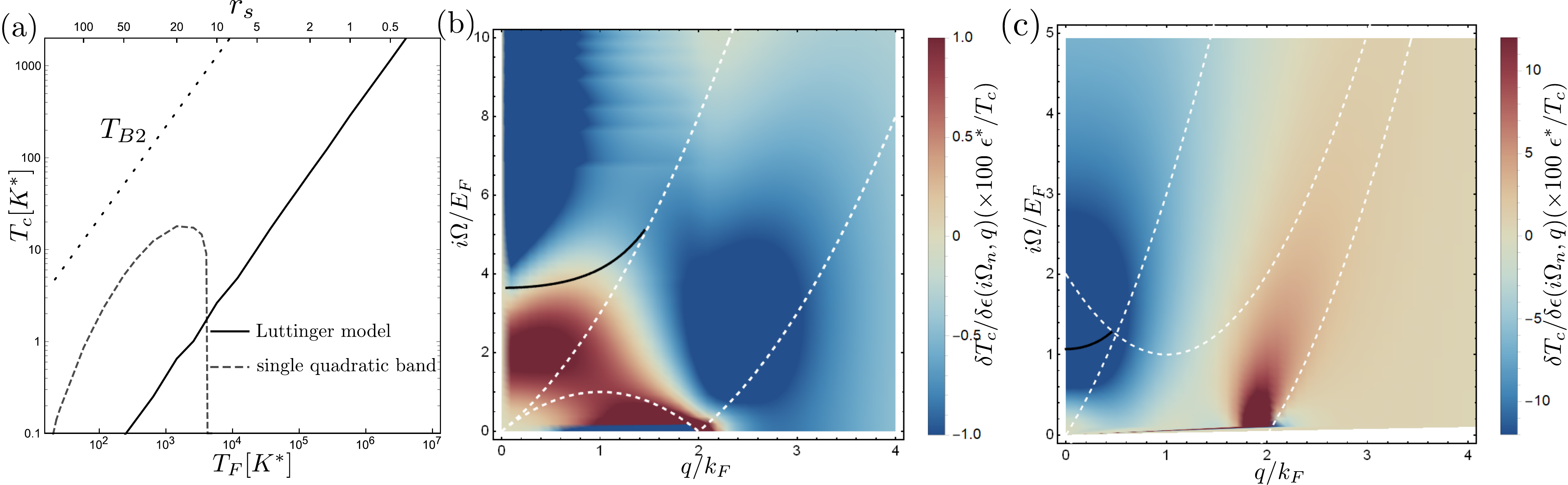}
    \caption{(a) Critical temperature in units of K$^* = m/(m_e\epsilon^{*2})$K for a single quadratic band (gray, dashed) and a Luttinger semimetal (plain, black). For comparison we superimpose the Bose-Einstein condensation temperature $T_{B2}$ for a density $n/2$ and a mass $2m$. Reproduced with permision from \cite{scus}. (b-c) The functional derivative ${\delta T_c}/{\delta \epsilon(i\Omega_n,q)}$ in percent of the critical temperature for (b) a single quadratic band structure and (c) a Luttinger semimetal, for $r_s = 15$. The white dashed lines are the branches of the particle-hole excitation diagram in real frequency and the black line is the plasmon dispersion in real frequency. The critical temperature is mostly sensitive to the dielectric function in the region of plasmons and near the $q = 2k_F$ static screening.}
    \label{fig:sols}
\end{figure}

We use this equation to reproduce the critical temperature for singlet $s-$wave pairing from the screened Coulomb repulsion in a single quadratic band structure \cite{takada2}, and compute it for a Luttinger semimetal (see Fig.~\ref{fig:sols}). For large Wigner-Seitz radii, the critical temperature of the Luttinger semimetal $T_c/T_F \approx 4.4\times 10^{-4}$ is smaller than for a single quadratic band, but extends to smaller values of $r_s$ \cite{scus}. We note that it was important to keep $\phi_+$ in Eq.~(\ref{eq:linelsh}), otherwise we would not find a solution. The value we obtain is comparable to the ratio $T_c/T_F \approx (2-5)\times 10^{-4}$ from measurements on the half-Heusler YPtBi \cite{yptbi1,yptbi2,yptbi3}. Because we have an $s-$wave superconductor, our result stands in contradiction with a recent proposition that YPtBi is a line-node superconductor~\cite{linenode} but this interpretation, based on magnetic properties, is arguable due to the small value of the lower critical field $B_{c1}$ in YPtBi \cite{bc1}, among other caveats. 

Because the critical temperature depends on an integral equation involving every component $(i\Omega_n, q)$ of the dielectric function $\epsilon(i\Omega_n, q)$, it is not straightforward to understand the origin of superconductivity. If one changes $\epsilon(i\Omega_n,q)$ by $\delta\epsilon(i\Omega_n,q)$ then the critical temperature $T_c$ changes by
\begin{equation}\label{eq:defdtc}
    \Delta T_c = 2\pi T\sum_{\Omega_n} \int dq~ \frac{\delta T_c}{\delta \epsilon(i\Omega_n,q)} \delta \epsilon(i\Omega_n,q) .
\end{equation}
The functional derivative ${\delta T_c}/{\delta \epsilon(i\Omega_n,q)}$ is a measure of the sensitivity of the critical temperature to screening. Here, the functional derivative can be decomposed into
\begin{equation}
    \label{eq:tride}
    \frac{\delta T_c}{\delta \epsilon(i\Omega_n,q)} = - \left.\frac{\delta  \rho_{}}{\delta \epsilon(i\Omega_n,q)}\right|_{T=T_c}\left/\frac{\partial  \rho_{}}{\partial T}\right|_{T=T_c}.
\end{equation}
In this equation, $\rho$ is the maximal eigenvalue of the linear Eliashberg equation~(\ref{eq:elshsym}). We use it to evaluate numerically the derivative $\partial\rho/\partial T|_{T=T_c}$ and we use the Hellmann-Feynman theorem to compute $\delta \rho/{\delta \epsilon(i\Omega_n,q)}$~\cite{scus}. In Fig.~\ref{fig:sols}(b,c), we show the sensitivity of the critical temperature $T_c$ to the different components of the dielectric function $\epsilon(i\Omega_n,q)$ for a quadratic band and a Luttinger semimetal~\cite{scus}. We notice larger values in the area associated to plasmons and close to $2k_F$, which are respectively associated to plasmon and Kohn-Luttinger mechanisms of superconductivity~\cite{takada,kohnluttinger}.

\section{Pairing beyond $s-$wave from spin-orbit coupling}

In Luttinger semimetals, the quasiparticles are described with $j = 3/2$ multiplets instead of spin-$1/2$ as in ordinary metals. The rotational symmetry of Eqs.~(\ref{eq:h0} - \ref{eq:hint}) allows to describe Cooper pairs by a gap function $\Delta^{J,LS}(i\omega_n,k)$ with a well defined total angular momentum $J$ that combines the pseudo-spin $S$ of the Cooper pair and its orbital angular momentum $L$ \cite{savary,ghorashi}. At the critical temperature, these gap functions satisfy the linear Eliashberg equations
\begin{align}
  \label{eq:djls}
  &\lambda(T) \Delta^{J,L_1S_1}_{\sigma_1}(i\omega_{n_1},k_1) = \\
  &- T \sum_{\substack{\ell\sigma_2\omega_{n_2}\\L_2S_2}} \int \frac{dk_2 k_2}{k_1} \frac{V_{\ell}(i(\omega_{n_1}-\omega_{n_2}),k_1,k_2) A^{J,L_1S_1L_2S_2}_{\ell,\sigma_1\sigma_2}(k_1, k_2)}{(\omega_{n_2}Z_{\sigma_2}(i\omega_{n_2},k_2))^2 + (\xi_{\sigma_2}(k_2) + \chi_{\sigma_2}(i\omega_{n_2},k_2))^2} \Delta^{J,L_2S_2}_{\sigma_2}(i\omega_{n_2},k_2),\nonumber
\end{align}
with $\lambda(T) = 1$ for $T = T_c$. Note that we have written the Eliashberg equation in its non-symmetrized form, in contrast to Eq.~(\ref{eq:elshsym}). This self-consistent relation can be complemented with off-diagonal components of the gap function~\cite{unconvsc}, that we neglect in the present discussion. In Eq.~(\ref{eq:djls}), the electron pairing is determined by $V_{\ell}$, the projection of the screened Coulomb potential $V_{0}({q})/\epsilon(i\Omega_n,{\bf q})$ on the $\ell-$th Legendre polynomial $P_{\ell}$, and by the form factor due to spin-orbit coupling :
\begin{equation}
  A^{J,L_1S_1L_2S_2}_{\ell,\sigma_1\sigma_2} = \frac{2\ell + 1}{2}\int \frac{ d^2{\bm \Omega}_{1}d^2{\bm \Omega}_{2} }{(2\pi)^3} P_{\ell}(\hat{\bf k}_1\cdot\hat{\bf k}_2) {\rm Tr}[ \hat{P}_{\sigma_1}({\bf k}_1) \hat{N}^{J,L_1S_1}({\bf k}_1) \hat{P}_{\sigma_2}({\bf k}_2) \hat{N}^{J,L_2S_2\dagger}({\bf k}_2)],
  \label{eq:ajls}
\end{equation}
where ${\bm \Omega}_{i}$ is the solid angle of ${\bf k}_i$. Here, the matrices $\hat{N}^{J,LS}({\bf k})$ correspond to the representation of the rotation symmetry on ${\bf J} = {\bf L} + {\bf S}$, 
\begin{equation}
  \hat{N}^{J,LS}({\bf k}) = \sum_{m_{L} m_{S}} C^{J}_{Lm_{L},S m_{S}} Y_{L m_{L}}(\theta_{\bf k},\phi_{\bf k}) \hat{M}_{S m_S} 
\end{equation}
where $C^{J}_{Lm_{L},S m_{S}}$ are the Clebsch-Gordan coefficients, $Y_{Lm_{L}}$ the spherical harmonics and $\hat{M}_{Sm_S}$ the pairing matrices with pseudo-spin $S$ of the Cooper pairs. Some of these combinations, for $L = 0,1$, are listed in \cite{savary,ghorashi}. We introduce the projectors $\hat{P}_{\pm}$ in Eq.~(\ref{eq:ajls}) to decompose the gap equation on the eigenstates of $\hat{H}_0$ on the upper ($+$) and lower ($-$) bands, that we respectively associate to eigenstates $\pm 3/2$ and $\pm 1/2$ of the helicity operator $\hat{\lambda} = \hat{\bf k}\cdot \hat{\bf J}$. Note that, for a given value of $J$, there is a finite number of components $V_{\ell}$ that contribute in the summation in Eq.~\eqref{eq:djls}. For example, for $s-$wave $J = L = S = 0$ only $\ell = 0$ and $2$ contribute. In the following, we further simplify Eq.~(\ref{eq:djls}) by considering a gap function in a unique $(J,L,S)$ sector, $\Delta^{J,LS}$, and write $A^{J,LSLS}_{\ell,\sigma_1\sigma_2} = A^{J,LS}_{\ell,\sigma_1\sigma_2}$. A more refined analysis would allow for mixing between different values of $(L,S)$ for a fixed $J$.

It is expected that the amplitude of the pairing potential depends on the largest combination of the coefficients $V_{\ell}$ and $A^{J,LS}$ close the Fermi surface, where $\sigma_1 = \sigma_2 = -$ and $k_1 = k_2 = k_F$. To be more accurate, one should consider the full $k-$dependence but let us work in this simpler limit. It was shown that this amplitude is the strongest for $J = L = S = 0$~\cite{savary},
which is precisely the order parameter we consider in our work (see section \ref{sec:sc}). This logic of maximizing the product $V_{\ell}A^{J,LS}_{\ell}$ applies well for superconductiviy from an attractive potential, like the electron-phonon coupling, where the eigenvalue of Eq.~(\ref{eq:djls}) with the largest absolute value, $\lambda_1(T)$, is already positive. However, it is not straightforward to extend to superconductivity from a repulsive potential, such as the Coulomb repulsion between electrons, where in the $s-$wave channel the eigenvalue with the largest absolute value is negative, $\lambda_1(T) < 0$, because of the overall repulsive nature of the Coulomb potential. Then, the $s-$wave solution to Eq.(\ref{eq:djls}) comes from the second largest-in-absolute-value eigenvalue, $\lambda_2(T) > 0$, which corresponds to the first electronic configuration where the Coulomb potential is attractive \cite{takada2}. 

Naively, the interband coupling may seem unfavoured due to the difference in energy between the two bands but it is worth considering its contribution. Indeed, we find that for $J = L = S = 1$, \emph{i.e.} $\hat{N}^{111} = \sqrt{3}(-{k}_z(\hat{J}_x + i \hat{J}_y) + (k_x+ik_y)J_z)/(\sqrt{5} k)$~\cite{savary}, the coupling is non-zero only for $\ell = 1$ and decomposes as a matrix on the bands with helicity $\pm 1/2$ and $\pm 3/2$ :
\begin{equation}
  A^{111}_{1} = \left(
    \begin{array}{cc}
      2/5 & 3/10\\
      3/10 & 0
    \end{array}
  \right).
\end{equation}
This matrix has eigenvalues $(2\pm\sqrt{13})/10$. Interestingly, one of them is negative~($\approx -0.16$) and thus, for the corresponding configuration of the gap function on the upper and lower bands, the eigenvalue $\lambda_1(T)$ with the largest absolute value can be positive instead of negative. This way, due to the difference in magnitude between $\lambda_1(T)$ and $\lambda_2(T)$, it may be possible to obtain a larger critical temperature with $J = L = S = 1$ (where $|\lambda_1| > |\lambda_2|$ and $\lambda_1 > 0$) than for $J = L = S = 0$ (where $|\lambda_1| > |\lambda_2|$ but $\lambda_1 < 0$) . However, for this to happen, the amplitude of the Coulomb repulsion has to overcome the difference in energy between the two bands and a more refined study is needed to evaluate the corresponding critical temperature.

\section{Conclusion}

Over a wide range of doping, we find that the $s-$wave critical temperature for a Luttinger semimetal with screened Coulomb repulsion is $T_c/T_F \approx 4.4\times 10^{-4}$. $T_c/T_F$ is small but may be an explanation for the superconductivity of YPtBi, a candidate Luttinger semimetal, where experiments report $T_c/T_F \approx (1-8)\times 10^{-4}$. Previous theoretical works on YPtBi, with phonon-based pairing, estimate a critical temperature at least one order of magnitude smaller than in experiments \cite{savary,meinert}. We quantitatively show the origins of superconductivity, in relation to the plasmon \cite{takada2} and Kohn-Luttinger \cite{kohnluttinger} mechanisms of superconductivity. We also analyze the Eliashberg equation of $j = 3/2$ fermions \cite{savary,ghorashi} and propose that an unconventional order parameter, with $J=L=S=1$, may turn the repulsive contribution of the screened Coulomb potential to attractive. This reminds a recent discussion on graphene, where the Berry curvature promotes the $\ell = 1$ component of a repulsive interaction to attractive~\cite{tommyli}. A more involved study would be required to determine the dominant superconducting channel.

\paragraph{Acknowledgments.}
This project is funded by a grant from Fondation Courtois, a Discovery Grant from NSERC, a Canada Research Chair, and a ``\'Etablissement de nouveaux chercheurs et de nouvelles chercheuses universitaires'' grant from FRQNT. This research was enabled in part by support provided by Calcul Qu\'ebec (www.calculquebec.ca) and Compute Canada (www.computecanada.ca).

%
%

\end{document}